\documentclass[preprint,review,12pt]{elsarticle}

\usepackage{graphicx}

\usepackage{amssymb}

\usepackage{pbox}
\usepackage{makecell}
\usepackage{hyphenat}

 \usepackage[top=2.5cm,left=4cm,right=4cm,bottom=2.5cm]{geometry}
\usepackage{verbatim} 
\usepackage{csvsimple}
\usepackage{multirow}
\usepackage{lipsum}
\usepackage{afterpage}

\usepackage{etoolbox}
\robustify\bfseries

\usepackage{eurosym}
\usepackage{siunitx}

    \DeclareSIUnit\eur{\officialeuro}
    \DeclareSIUnit\M{M}
    \DeclareSIUnit\k{k}

  \def\sym#1{\ifmmode^{#1}\else\(^{#1}\)\fi}

\usepackage{setspace} 

\usepackage{csquotes}
\usepackage{amsmath}
\usepackage{bm}

\usepackage{xspace}
	\newcommand\ie{i.\,e.\xspace}
	\newcommand\eg{e.\,g.\xspace}

	\newcommand\US{US\xspace}

\usepackage[amsmath,hyperref,framed]{ntheorem} 
  \theoremstyle{plain}
     
  \theoremstyle{nonumberplain}
    \theoremseparator{.}
    \theoremheaderfont{\bfseries}
    \theorembodyfont{\normalfont}
    \theoremsymbol{$\blacksquare$}
      \RequirePackage{amssymb}

      \makeatletter
    \let\copy@theorem@headerfont=\theorem@headerfont
    \newcommand{\my@theorem@headerfont}{%
        \boldmath\copy@theorem@headerfont\unboldmath
      }
    \let\theorem@headerfont=\my@theorem@headerfont
      \makeatother

\theoremstyle{nonumberplain}
\theoremseparator{.}
\usepackage[%
  sort&compress
]{cleveref}
\usepackage{url}

\usepackage{isomath}

\usepackage[usenames,dvipsnames]{xcolor}

\usepackage{colortbl}
\usepackage{tabularx}
\usepackage{booktabs}
\usepackage{collcell}
\usepackage{longtable}

\usepackage{ragged2e}
  
\newcommand{\PreserveBackslash}[1]{\let\temp=\\#1\let\\=\temp}
\newcolumntype{v}[1]{>{\PreserveBackslash\RaggedRight\hspace{0pt}}p{#1}}

  \usepackage{adjustbox}
\newcolumntype{Q}[2]{%
    >{\adjustbox{angle=#1,lap=\width-(#2)}\bgroup}%
    l%
    <{\egroup}%
}
\newcolumntype{L}[1]{>{\raggedright\let\newline\\\arraybackslash\hspace{0pt}}m{#1}}
\newcolumntype{C}[1]{>{\centering\let\newline\\\arraybackslash\hspace{0pt}}m{#1}}
\newcolumntype{R}[1]{>{\raggedleft\let\newline\\\arraybackslash\hspace{0pt}}m{#1}}

\usepackage[
    colorinlistoftodos,
    textsize=footnotesize,
        ]{todonotes}

\makeatletter
    \renewcommand{\fps@figure}{htb}         
    \renewcommand{\fps@table}{htbp}         
\makeatother 

\usepackage{float}
\usepackage{rotating}

\usepackage{placeins}

\usepackage{algpseudocode} 
\usepackage{algorithm}

\biboptions{authoryear}

\journal{arXiv}

\begin{document}

\begin{frontmatter}



\title{Public decision support for low population density areas: An imbalance-aware hyper-ensemble for spatio-temporal crime prediction}

\author[ETHaddress]{Cristina Kadar\corref{mycorrespondingauthor}\fnref{equal}}
\fntext[equal]{These authors contributed equally to this work and are listed alphabetically.}
\cortext[mycorrespondingauthor]{Corresponding author.}
\ead{ckadar@ethz.ch}

\author[ETHaddress]{Rudolf Maculan\fnref{equal}}
\ead{rudolfmaculan@me.com}

\author[ETHaddress]{Stefan Feuerriegel}
\address[ETHaddress]{ETH Zurich, Weinbergstr. 56/58, 8092 Zurich, Switzerland}
\ead{sfeuerriegel@ethz.ch}

\begin{abstract}
Crime events are known to reveal spatio-temporal patterns, which can be used for predictive modeling and subsequent decision support. While the focus has hitherto been placed on areas with high population density, we address the challenging undertaking of predicting crime hotspots in regions with low population densities and highly unequally-distributed crime.This results in a severe sparsity (\ie, class imbalance) of the outcome variable, which impedes predictive modeling. To alleviate this, we develop machine learning models for spatio-temporal prediction that are specifically adjusted for an imbalanced distribution of the class labels and test them in an actual setting with state-of-the-art predictors (\ie, socio-economic, geographical, temporal, meteorological, and crime variables in fine resolution). The proposed imbalance-aware hyper-ensemble increases the hit ratio considerably from \SI{18.1}{\percent} to \SI{24.6}{\percent} when aiming for the top \SI{5}{\percent} of hotspots, and from \SI{53.1}{\percent} to \SI{60.4}{\percent} when aiming for the top \SI{20}{\percent} of hotspots.
As direct implications, the findings help decision-makers in law enforcement and contribute to public decision support in low population density regions.
\end{abstract}


\end{frontmatter}

\section{Introduction}
\label{sec:introduction}

Crime inflicts immense financial losses upon individuals, businesses, and organizations, and can even threaten the stability of societies. For instance, according to recent figures from the Federal Bureau of Investigation, annual financial losses due to property crime in the United States amount to 3.6~bn USD, with an average cost of 2,361 USD per burglary.\footnote{US Federal Bureau of Investigation, Uniform crime report: \url{https://ucr.fbi.gov/crime-in-the-u.s/2016/crime-in-the-u.s.-2016/topic-pages/burglary}. Last accessed: March~11, 2018.} Beyond the financial damage, crime incidents are also known to trigger negative social and psychological effects, since victims suffer from a heightened level of perceived risk, which has been found to result in a significant decrease in the quality of life \citep{Doran2012}.
Hence, it is the objective of decision-makers in the private and public sectors to find strategies for effective crime prevention. 

In the effort to reduce crime, governments and law enforcement agencies, such as US police departments, have recently started experimenting with techniques for predictive policing in order to optimize the use of resources and to increase the chances of deterring, as well as preventing, crime events.\footnote{US Federal Bureau of Investigation, Articles: \url{https://leb.fbi.gov/articles/featured-articles/predictive-policing-using-technology-to-reduce-crime}. Last accessed: August~4, 2018.} The term \textquote{predictive policing} refers to the use of predictive analytics with the aim of identifying the potential location of criminal activity prior to such an event taking place \citep{Ratcliffe2014}. 
Formally, this approach draws upon historical records of crime events in order to make spatio-temporal forecasts \citep{Bowers2004, Mohler2011}.\footnote{We point out that, throughout this work, we build upon the place-centric notion of crime prediction, with the aim of forecasting time-dependent spatial hot spots of elevated crime risk. This is in contrast to a people-centric notion, such as adopted by \citet{Canter2000} or \citet{Wang2013}, which aims at identifying attributes of potential offenders.} In addition, the predictive models are often extended by further information related to the socio-economic status of the resident population and to nearby points of interest (POI) \citep{Kadar2017, Rummens2017, Vomfell2018, Wang2012, Xue2006}, basic temporal variables \citep{Rummens2017, Wang2012}, or even social media, telecom, or mobility data \citep{Bogomolov2014, Gerber2014, Kadar2018, Vomfell2018, Wang2016}, in order to better adapt to the spatio-temporal nature of crime events. 
 
\begin{figure}[H]
\centering
\includegraphics[width=0.9\textwidth]{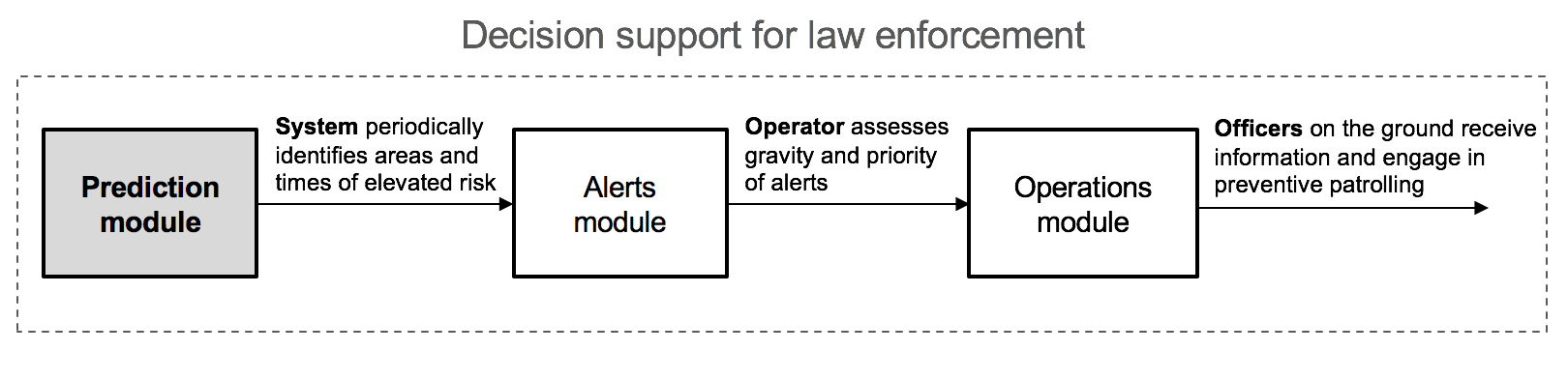}
\caption{Schematic illustration of how predictive policing delivers spatial decision support for law enforcement.}
\label{fig:decision_support}
\end{figure}

Forecasts from predictive policing improve situational awareness at both the tactical and strategic levels for law enforcement bodies and help them develop strategies for more efficient and effective policing \cite[p.~2]{Perry2013}. \Cref{fig:decision_support} summarizes the main steps involved in deriving tactical decision support from predictive policing. In doing so, predictive policing is based on the assumption that the presence of police offers at crime hotspots leads to decreasing crime rates, which has been recently validated in randomized controlled trials \citep{Mohler2015}.

Previous research has developed models for crime prediction that target highly populated areas. Examples include cities such as Los Angeles \citep{Gerber2014}, London \citep{Bogomolov2014}, and Liverpool \citep{Bowers2004}. Other studies even narrow the focus to individual districts such as the San Fernando Valley in Los Angeles \citep{Mohler2011}. Yet there is scant evidence that predictive policing can also be applied to areas with lower population density. 
In fact, prior literature has overlooked sparsely-populated regions, despite the fact that over 50~percent of household burglaries in the \US occur in such areas.\footnote{Bureau of Justice, National crime victimization survey: \url{https://www.bjs.gov/index.cfm?ty=nvat}. Last accessed: March~11, 2018.} 
 However, this segment of society is currently not benefiting from novel, data-driven techniques for public decision support.

The key contribution of this work is to adapt predictive policing to areas with low population density. The unique features of these regions require extensive modifications to current models used in predictive policing. More specifically, are characterized by low population densities and crime incidents that are distributed sparsely. 
 In fact, only \SI{0.06}{\percent} of the total daily observations in our study reflect a crime event. As a consequence, the outcome variable is affected by a severe sparsity, which, in machine learning, is called class imbalance. Due to it, traditional approaches to predictive modeling struggle achieving a forecast performance beyond a random vote. As a remedy, we follow recent suggestions for handling class imbalances, 
 and develop a hyper-ensemble for spatio-temporal crime prediction that is specifically suited to an extreme class imbalance and, thus, to low population density areas.

Our evaluation demonstrates the capacity of crime prediction in a real-world, low population density setting. Our results reveal the challenge of forecasting spatio-temporal crime patterns with na{\"i}ve predictive models, since these outperform the default hit rate of a majority vote by a mere $18.1$ percentage points in identifying the top \SI{5}{\percent} of crime hotspots. To improve the predictive power, we propose a hyper-ensemble that combines the benefits of under-sampling and ensemble learning, thereby modeling decisive relationships between predictors and outcomes even in the presence of sparse crime events and thus extreme class imbalances. As a result, our hyper-ensemble consistently yields considerable performance improvements over common baselines: it increases the hit ratio significantly from \SI{18.1}{\percent} to \SI{24.6}{\percent} when aiming for the top \SI{5}{\percent} of hotspots, and from \SI{53.1}{\percent} to \SI{60.4}{\percent} when aiming for the top \SI{20}{\percent} of hotspots.

Our work entails immediate implications for decision support, especially across the public sector. 
This manuscript helps to further develop decision-making in public bodies by incorporating spatial analytics for data-driven decision support. Furthermore, literature commonly studies decision support in high population density regions, while neglecting a major share of the population that lives in areas with lower population density. Here we provide specific levers for translating existing prediction algorithms, such as those used for managing rescue units or traffic flow, to these settings. This is a direct remedy for an acute societal challenge, since sparsely-populated areas already experience lower average incomes and are now additionally excluded from the potential benefits of more efficient decision-making. 

The remainder of this paper is structured as follows. \Cref{sec:related_work} reviews theoretical and empirical efforts concerning crime prediction, thereby revealing the dearth of evidence in low population density environments. To close this gap, \Cref{sec:methods} proposes our hyper-ensemble for crime prediction in the case of extreme class imbalance. Its performance is evaluated in \Cref{sec:results}, revealing considerable improvements over traditional predictive models. \Cref{sec:discussion} discusses our findings in the context of managerial implications and public decision support, while \Cref{sec:conclusion} concludes.   

\section{Related work}
\label{sec:related_work}

This section provides a detailed overview of the theoretical foundations, drawn from the field of criminology, based on which we motivate common choices in predictive modeling of crime incidents. 

\subsection{Theoretical foundation}

The spatial nature of crime has been subject to extensive theory development. In this regard, under the umbrella of \emph{crime pattern theory}, individual locations have been categorized according to whether they act as crime generators, crime attractors or crime detractors \citep{Brantingham1995}. For instance, locations where large crowds assemble are supposed to serve as crime generators (\eg, sporting events), while the intrinsic characteristics of others function as crime attractors (\eg, bars) or crime detractors (\eg, police stations). In practice, these patterns can be modeled by the inclusion of points-of-interests~(POIs) and other infrastructure characteristics as factors in predictive modeling, an approach that we also follow in our work. In addition, the \emph{social disorganization theory} \citep{Shaw1942} and its further offshoots link crime levels to the ecological attributes of the neighborhood such as socio-economic status, residential stability, and ethnic diversity. This motivates our choice of predictors in order to account for socio-demographic and economic variations among the resident population.

The temporal nature of crime is often theorized to follow two distinct patterns \citep{FarrellG.1993}. On the one hand, the concept of \emph{repeat victimization} proposes that crime events are more likely to occur at locations at which other crime incidents have previously taken place. The reason for the increased risk level originates from the assumption that offenders are more likely to exploit suitable opportunities further, for example, by stealing objects replaced after the initial theft. On the other hand, \emph{near repeat victimization} refers to crime events occurring to close proximity of locations of past incidents. Here theory assumes the concept of risk heterogeneity, which states that the only association between one offense and another is the target involved. Since nearby locations of an existing crime scene are more likely to share certain characteristics, such as escape routes or levels of surveillance, it renders them potential locations for further crime in the short run. Theoretical arguments have been proposed for both patterns \citep{Johnson2008}, and we thus take these theories into consideration by incorporating counts of previous crime incidents into our predictive models.

Finally, the characteristics of an environment can inherently change according to climatic and seasonal conditions. A detailed literature review of different studies concerning the impact of weather-related variables on crime was performed in \citet{Murataya2013}. The fact that both violent and property crimes are significantly correlated with major holidays is documented in \citet{Cohn2003}. These studies have informed our choice of further temporal factors.


\subsection{Crime prediction}

\subsubsection{Na{\"i}ve predictions from historic crime data}

Early attempts to identify crime hotspots relied upon non-parametric approaches, thus benefiting from simple estimation procedures but neglecting the prognostic capacity of environmental attributes and all associated spatio-temporal dynamics. For instance, the so-called spatial hotspot model applies a simple kernel density estimation to historic crime events in order to locate areas that were previously associated with a higher likelihood of criminal activity \citep{Chainey2008}. While this approach proved feasible in high population density settings, the disparate and sparse crime events in less populated areas limit its applicability. Nevertheless, historic crime data serves as one of our baselines for determining locations with a high risk of crime. In fact, our empirical results later establish that basic models without theory-informed crime correlates result in inferior performance as compared to models leveraging spatio-temporal predictors.

\subsubsection{Machine learning models with spatio-temporal predictors}

Machine learning allows for the incorporation of crime correlates in order to improve prediction performance. It can thereby accommodate further theories, such as crime pattern theory and social disorganization theory. As a result, a variety of models and predictors have been proposed in the literature, which we summarize in the following (see \Cref{tab:review}).

There is considerable variability in terms of model choice. Past studies have quantified the probability of criminal events by means of generalized additive models \citep{Wang2012}, logistic regression \citep{Gerber2014,Rummens2017}, gradient boosting \citep{Vomfell2018}, neural networks \citep{Rummens2017}, or random forests \citep{Bogomolov2014,Vomfell2018}. However, there is no evidence that any one model is consistently superior to all others. A potential reason might be located in the different prediction horizons, which can vary from month-ahead predictions \citep{Bogomolov2014,Wang2012} and bi-weekly crime counts \citep{Vomfell2018} to ranking hotspots on a daily basis \citep{Gerber2014}. Notably, these works all deal with urban data and thus avoid having to account for class imbalance. Hence, we later experiment with a wide range of models in order to identify a tailored prediction strategy for our research setting.  

Numerous spatio-temporal crime correlates have been used as predictors, often in a theory-informed manner. Location features inspired by social disorganization theory and crime pattern theory are common and predominantly include socio-demographic variables, infrastructure and POI data \citep{Bogomolov2014,Kadar2017,Rummens2017,Vomfell2018,Wang2012}. To account for (near) repeat victimization, previous crime has been incorporated into the prediction models \citep{Gerber2014, Wang2012}. Further dynamic features refer to seasonal indicators \citep{Rummens2017}, or urban human dynamics extracted from social media, mobility, or telecom data \citep{Bogomolov2014,Gerber2014,Kadar2018,Vomfell2018}. 
We adhere to these works and follow an extensive, theory-informed selection of spatio-temporal predictors in our low population setup.


\Cref{tab:review} summarizes key studies on short-term crime prediction from the literature. We note that all studies restrict the analysis to an area with high population density (a major city or a region of it) and do not consider areas with low population density. Therefore, the novelty of this work is to expand crime prediction to low population density settings, which necessitates our hyper-ensemble, since it can successfully handle extremely imbalanced distributions of crime events. 

\begin{sidewaystable}
\begin{table}[H]
\centering
\scalebox{0.65}[0.65]{
\begin{tabular}{lp{3.5cm}l>{\raggedright\arraybackslash}p{6cm}lp{4.8cm}>{\raggedright\arraybackslash}p{6cm}}
\toprule	
\textbf{\footnotesize{}Study} &\textbf{\footnotesize{}Population density} &\textbf{\footnotesize{}Study area} &\textbf{\footnotesize{}Spatial resolution} &\textbf{\footnotesize{}Temporal resolution} &\textbf{\footnotesize{}Features} &\textbf{\footnotesize{}Method}\\ 
\midrule
\citet{Bowers2004}     &High        &Liverpool                 &Grid cells of 50\,m $\times$ 50\,m 
&2 days / 1 week       &Crime                                         &Prospective hotspot\\[20pt] 
\citet{Mohler2011}     &High  &Part of Los Angeles       &Grid cells of 200\,m $\times$  200\,m        
&1 day                 &Crime                                         &Self-exciting point processes\\[20pt] 
\citet{Wang2012}       &High        &Charlottesville           &Grid cells of 32\,m $\times$  32\,m           
&1 month               &Crime, spatial                    & \nohyphens{Spatio-temporal generalized additive model}\\[20pt] 
\citet{Gerber2014}     &High   &Los Angeles               &Points at 200\,m $\times$  200\,m intervals  
&1 day                 &Crime, social media                          &Logistic regression\\[20pt]
\citet{Bogomolov2014}  &High   &London metropolitan area  &Lower layer super output areas (geographic hierarchy)           
&1 month               &Crime, spatial, telecom                     &Random forest\\[20pt]
\citet{Rummens2017}    &High        &Unnamed city in Belgium   &Grid cells of 200\,m $\times$  200\,m         
&2 weeks               &Crime, spatial, temporal                               &Logistic regression, neural network\\[20pt]
\citet{Vomfell2018}    &High  &New York City             &\pbox{6cm}{Census tracts \\(geographic hierarchy)}         
&1 week                &Crime, spatial, social media, mobility     &Random forest, gradient boosting, neural network\\[20pt]
\midrule
This work              &Low  &Swiss canton Aargau (analogous to an US state)  &Grid cells of 200\,m $\times$  200\,m         
&1 day                 &Crime, spatial, temporal               &Hyper-ensemble\\[20pt] 
\bottomrule
\end{tabular}
}
\caption{Key studies on short-term crime prediction (\ie, up to one month as required in tactical setups). The overview shows the dearth of works that target low population density areas. Aargau has a total surface of 140,400 hectares (1.8 times bigger than New York City) and a population density of 4.72 people/hectare (seven times less populated than New York City.}
\label{tab:review}
\end{table}
\end{sidewaystable}

\section{Methods and materials}
\label{sec:methods}

\subsection{Research framework}
\label{sec:approach}

Low population density areas are characterized by a strong sparseness of crimes relative to the total number of instances. Such a setting where the predicted variable in machine learning is subject to sparse outcomes is termed \emph{class imbalance}. That is, the number of instances from one class outnumbers the number of instances from the other class. In our particular setting, the dependent variable will be zero ($=$ no crime) almost everywhere and only set to one ($=$ crime) in $< \SI{0.1}{\percent}$ of cases. This poses a great challenge for identifying the few crime events through common predictive modeling and necessitates a tailored approach to handle such severe class imbalance. 

Only a few studies in the decision support literature have investigated prediction in imbalanced datasets, either as a general methodology \citep{Piri2018} or for specific application domains such as bankruptcy prediction \citep{Veganzones2018}. However, we later see that the traditional approach of over- or under-sampling is not sufficient in the case of such an extreme imbalance and, for this reason, we propose an imbalance-aware hyper-ensemble, which performs \emph{repeated} under-sampling and should thus be more robust in identifying decisive relationships for predictive purposes. 

\begin{figure}[H]
\centerline{\includegraphics[width=0.9\textwidth]{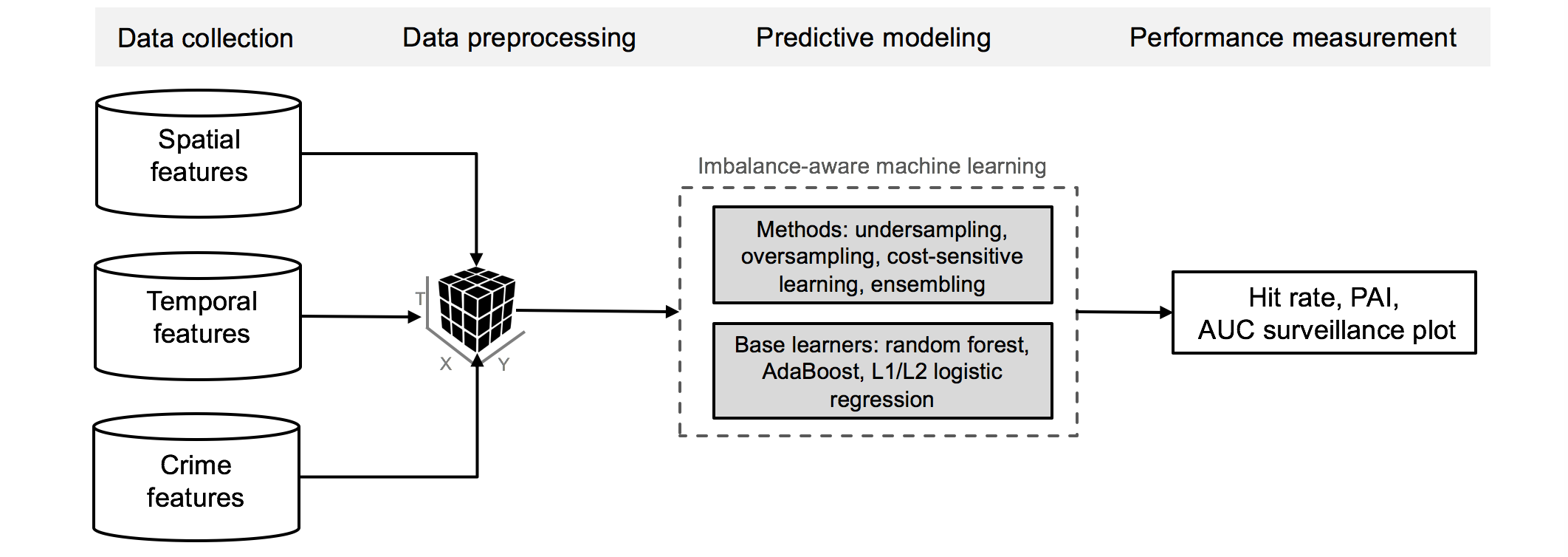}}
\caption{{\small{}Research framework combining different feature sets in order to generate forecasts of crime hotspots based on imbalance-aware machine learning.}}
\label{fig:research_evaluation}
\end{figure}

\Cref{fig:research_evaluation} summarizes our overarching research framework for identifying the most likely times and locations of burglary in a low population density environment. We investigate our proposed hyper-ensemble, but also experiment with various resampling strategies and cost-sensitive models as baselines; see the specification in the subsequent sections. 

\subsection{Methods for imbalance-aware machine learning}
\subsubsection{Proposed hyper-ensemble}

Our proposed hyper-ensemble is intended to address the skewed distribution towards one class in the data and the fact that classical over- and under-sampling approaches are likely to miss discriminative information in the data. To address this and give the learning procedure additional structure, our hyper-ensemble is trained on different subsamples of the data and the test set results of these models are averaged in a final single prediction, as detailed in \Cref{alg:gsp}.


\begin{algorithm}
  \caption{Hyper-ensemble of random under-sampling models.}
\footnotesize
  \textsc{Training phase}
  \begin{algorithmic}[1]
	  \Require Given training set $D_\mathrm{train}$ and the number of elements in the ensemble ($\phi$)
    \For {$i = 1$ to $\phi$}:
    		\State Create balanced sub-sample $D_\mathrm{train}^i$  of $D_\mathrm{train}$ by randomly drawing instances without replacement from the majority class until the same number of instances as in the minority class is reached
		\State Train model $M^i$ with sub-sample $D_\mathrm{train}^i$ 
    \EndFor 
		\item \textbf{return} ensemble $\{ M^1, \ldots, M^\phi \}$
  \end{algorithmic}
			
	\vspace{0.3cm}
	\textsc{Prediction phase}
	 \begin{algorithmic}[1]
		\Require Given an unseen observation $\mathbold{x}$ and the ensemble $\{ M^1, \ldots, M^\phi \}$
    	 \For {$i = 1$ to $\phi$}:
      	 	\State  Apply model $M^i$ to $\mathbold{x}$ and obtain the predicted probability $\hat{y}^i$ 
    	 \EndFor 
			
    	\item \textbf{return} final probability whether $\mathbold{x}$ is labeled as crime or no-crime by averaging over the ensemble, \ie, $\hat{y} = \sum_{i=1}^{\phi}{\hat{y}^i}/\phi$
  \end{algorithmic}
  \label{alg:gsp}
\end{algorithm}

In this way, we combine the two concepts of under-sampling (\ie,  adjusting the class distribution of the data set) and ensemble learning (\ie, strategically generating and combining multiple models) towards a strategy that exploits more information than a standard under-sampling approach and provides additional structure in comparison to a standard over-sampling approach.


\subsubsection{Baselines}
We evaluate our approach against several baselines.
The most basic baseline is that of the majority class classifier -- in our case this classifier will always categorize a spatio-temporal unit as no-crime (majority class). The na{\"i}ve classifier baseline is that of a normal machine learning classifier, where no adjustments were made to account for the class imbalance. 

In addition, a cost-sensitive learning model is implemented by proportionally increasing the cost of classification mistakes of the minority class. This classifier is trained on the entire dataset, and is an instantiation of an algorithmic approach to handling class imbalances. 

We further experiment with resampling approaches to class imbalance. Random over-sampling achieves a balanced training dataset by randomly duplicating samples from the minority class until the minority class reaches the same number of samples as the majority class. In turn, random under-sampling yields a balanced training dataset by selecting a random subset of the majority class with the same size as the minority class. 
Heuristic over-sampling is implemented by means of SMOTE \citep{Chawla2002}, where, in order to over-sample the minority class to the size of the majority class, the following algorithm is applied: for each point $p$ in the minority class, choose a random point $r$ among its three nearest neighbors in the same class and create a random point on the line between $p$ and $r$.
Heuristic under-sampling is performed by applying the NearMiss method \citep{Yen2006}, where, in order to reach class balance, those points in the majority class are retained that are closest to their three nearest neighbors in the minority class.

\subsection{Estimation procedure}	
Our resampling strategies make no assumptions regarding the underling base learners, giving us the flexibility to experiment with different choices. We have evaluated different models\footnote{Due to their internal structure that automatically ranks the importance of features, these models are able to cope with datasets that contain many heterogeneous and potentially collinear features.}, including regularized linear models (logistic regression with L1 and L2 regularization, \ie, LASSO and ridge logistic regressions), bagging (random forests), and boosting (AdaBoost). They all return a probability score in a binary classification setup, which we can then sort and use to compute top hotspots. We implement all of the aforementioned base learners but, for the sake of brevity, we report exhaustively the results of only one classifier, namely the random forest classifier, since it returned best results across different specifications in our experiments. This is in line with prior evidence, as random forests are known to return best benchmarking results across a multitude of problems and metrics \citep{Caruana2006}. 

We briefly summarize the idea behind the base learners in the following. Random forests are an ensemble learning method for classification, where an entire set of decision trees are grown at training time, and their mode prediction is output at testing time, thus lowering the variance of the individual trees.
AdaBoost is also an ensemble model, but instead of averaging the prediction results of decision trees trained in parallel, it trains sequential decision trees, such that each new learner is optimized to correctly classify instances that have been misclassified by the previous learners. The L1 and L2 logistic regressors are simpler (\ie, linear) models with automatic regularization to avoid overfitting.

In all experiments, the dataset is split into $2/3$ training data (first two years) and $1/3$ testing data (third year). This split specifically maintains the chronological order of the data, simulating a real-world scenario and avoids over-optimistic inference \citep{Faraway2016, Hirsch1991}. Furthermore, the classifiers are trained in a 5-fold cross validation setup on the training set in which the optimal hyper-parameters are identified, and results are always reported on the test set. Optimized parameters consist of the number and depth of the decision trees for random forest, number of trees and learning rate for AdaBoost, and regularization strength for regularized logistic regressions.

\subsection{Prediction performance}


For predictive policing, a decision-maker is usually interested in how many of the top hotspots were correctly identified. In machine learning, this is referred to as a ranking task. Domain-specific metrics have been proposed in the literature to evaluate spatio-temporal prediction models. The hit rate metric is the percentage of crime cells within a specified time period falling into the areas where crimes are predicted to occur \citep{Chainey2008} and is defined as $\mathit{hit\ rate = n / N}$, where $n$ represents the number of crime areas correctly predicted, and $N$ refers to the total number of crime areas within the studied area.

The second metric with which to compare geographical prediction models is the prediction accuracy index (PAI) and has been utilized in multiple crime studies \citep{Adepeju2016,Chainey2008,Rummens2017}. It incorporates a trade-off between the hit rate of the identified hotspots and their relative size: $\mathit{PAI\ = hit\ rate\ / coverage\ area}$. The greater the number of future crime events in a hotspot area that is smaller in size compared to the whole studied area, the higher the PAI value. The metric is controlled by the coverage area parameter, which defines what percentage of the whole area the police would be able to patrol \citep{Adepeju2016} and is defined as $\mathit{coverage\ area = a / A}$, where $a$ is the combined area of all predicted hotspots and $A$ is the total studied area. The coverage area could be set by the decision-maker to, \eg, \SI{5}{\percent}. Therefore, in the results section, we present the metrics under different coverage area specifications. The predictions are made on a daily basis and averaged across all days in the test set.

Finally, a surveillance plot depicts the hit rate values $y$ as a function of increasing coverage area values $x$ \citep{Gerber2014}. The surveillance plot is a powerful tool: if one were to monitor the top $x$ most threatened areas according to the prediction, one would observe approximately $y$ of all crimes. To produce a scalar summary for surveillance curves, we calculate their total area under the curve (AUC) \citep{Gerber2014}.

\subsection{Data}

Our crime dataset consists of burglary incidents in the Swiss canton\footnote{Similar to an US state, a Swiss canton is a subdivision of a country established for political or administrative purposes. As of 2018, Switzerland comprises 26 cantons.} of Aargau (\ie, analogous to a State in the US) over the course of three years from January~14, 2014 to January~13, 2017. Switzerland has consistently ranked in past several years as one of the top destinations for burglars in Europe.\footnote{Eurostat, Recorded offences by offence category (police data): \url{http://ec.europa.eu/eurostat/web/products-datasets/-/crim_off_cat}. Last accessed: February~28, 2018.} 
{The canton of Aargau fits our definition of a low population area: given a total size of 140,400 hectares, it has a total population of approximately 660,000 inhabitants. Highest population densities (16 to 20 people/hectare) are achieved only in the cities of  Aarau (20,000 inhabitants), Baden (17,500 inhabitants), Brugg (10,000 inhabitants), and Zofingen (10,500 inhabitants), with the majority of the environment being sparsely populated. The overall population density amounts to only 4.72 people/hectare.\footnote{\url{https://www.ag.ch/en/weiteres/portrait/zahlen_und_fakten/zahlen_und_fakten.jsp}. Last accessed: January~17, 2019}. 

Each offense in the provided crime dataset features the exact geolocation, and the reported time period in which the crime was committed.
\Cref{fig:Crime-Heatmap} presents a heat map of all burglary incidents. We notice that burglary tends to cluster in three more densely populated areas, although incidents occur in less inhabited areas, as well. All burglary incidents span \SI{32.7}{\percent} of the cantonal built surface.
 
\begin{figure}[H]
\centering
\includegraphics[width=0.75\textwidth]{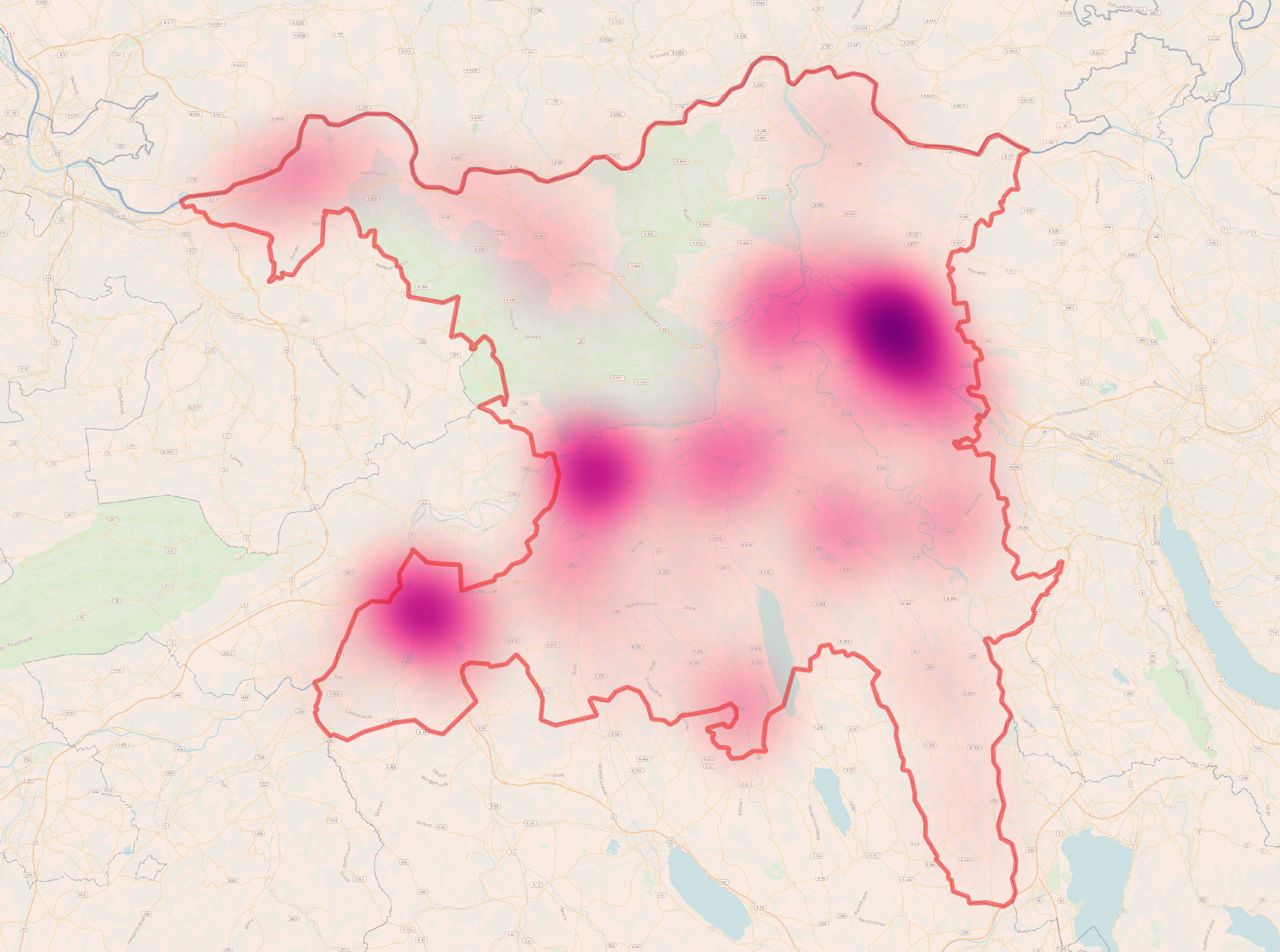}
\caption{{\small{}Heat map of burglary incidents over three years in our study setup.}}
\label{fig:Crime-Heatmap}
\end{figure}
 
We preprocess the data such that we only consider cells with built land-use where a burglary is theoretically possible, like residential and industrial areas, and exclude unbuilt land-use, like forests (see the online supplements for details). This filtering process is based on land use data provided by the cantonal geoportal AGIS\footnote{AGIS: \url{https://www.ag.ch/de/dfr/geoportal/geoportal.jsp} -- \textquote{Bauzonen Schweiz (harmonisiert), Ausschnitt AG, gemaess MGDM}. Last accessed: February~28, 2018.} and yields a final number of 10,149 grid cells that could be matched with a crime time series. Following this step, the average population density of the cells remains low at 15.6 people/hectare. This is, for instance, seven times smaller than the average population density in New York City. 

Spatio-temporal data was collected with a spatial resolution of either point-level or in discretized cells of one hectare (100\,m $\times$ 100\,m) from the original data sources. All spatio-temporal features were then aggregated to grid cells of 200\,m $\times$ 200\,m with daily resolution. This choice is consistent with leading studies in the literature \citep{Gerber2014,Mohler2011,Rummens2017} and was made jointly with the decision-maker, \ie, the police analyst. It was essential for the decision-maker that the spatial and temporal unit of analysis be kept at low granularity in order for the police to plan effective preventive actions. 
The final dataset consists of 11,123,304 spatio-temporal observations ($=$ 10,149 cells $\times$ 1,096 days), out of which only 6,266 observations are labeled as containing a burglary event. This amounts to only \SI{0.06}{\percent} of observations of the positive class, highlighting a very severe class imbalance in the dependent variable.

\begin{sidewaystable}
\centering
\scalebox{0.78}[0.78]{
\begin{tabular}{p{1cm}p{1cm}p{7cm}p{3cm}p{11cm}p{4cm}}
\toprule
\textbf{\footnotesize{}Type} &\textbf{\footnotesize{}Theory} &\textbf{\footnotesize{}Name} &\textbf{\footnotesize{}Dimension} &\textbf{\footnotesize{}Description} &\textbf{\footnotesize{}Source}
\tabularnewline
\midrule

\rotatebox[origin=c]{90}{\scriptsize{} \shortstack[l]{Crime}} &\rotatebox[origin=c]{90}{\scriptsize{} \shortstack[l]{(Near) \\ repeat \\ victimization}}
&\emph{ \raggedright{\scriptsize{}prior1d, prior3d, prior7d, prior14d}} & {\scriptsize{}integer} & \raggedright{}{\scriptsize{}Number of offenses in the respective and neighboring cells in the past days} & \multirow{1}{*}{\scriptsize{} \shortstack[l]{Aargau cantonal police}} \tabularnewline 
\midrule

\multirow{29}{*}{\rotatebox[origin=c]{90}{\scriptsize{} \shortstack[l]{Locational}}} &\multirow{15}{*}{\rotatebox[origin=c]{90}{\scriptsize{} \shortstack[l]{Social disorganization theory}}}
&\emph{\scriptsize{}popdens} & {\scriptsize{}people/hectare} & \raggedright{}{\scriptsize{}Density of total residential population} & \multirow{8}{*}{\scriptsize{} \shortstack[l]{AGIS: ``Statistik der \\ Bevoelkerung auf \\ Hektarbasis''}} \tabularnewline 
&&\emph{\scriptsize{}popbirth\_nonCH} & {\scriptsize{}percent} & \raggedright{}{\scriptsize{}Fraction of residents not born in Switzerland}\tabularnewline 
&&\emph{\scriptsize{}popcit\_EU, popcit\_europ, popcit\_noneurop, popcit\_CH} & {\scriptsize{}percent} & \raggedright{}{\scriptsize{}Fraction of residents: EU, non-EU European, non-European, or Swiss citizens}\tabularnewline 
&&\emph{\scriptsize{}popcit\_dividx} & {\scriptsize{}real between 0 and 1} & \raggedright{}{\scriptsize{}Diversity\footnote{Diversity is computed as the normalized Shannon entropy, given all possible categories for the variable, similarly to the approach in \citet{Kadar2018}. The resulting value lies between 0 (homogenous) and 1 (diverse).} of citizenship.  }\tabularnewline 
&&\emph{\scriptsize{}pop\_age1, pop\_age2, pop\_age3, pop\_age4} & {\scriptsize{}percent} & \raggedright{}{\scriptsize{}Fraction of residents between 0-19, 20-34, 35-64 and 65+ years of age }\tabularnewline 
&&\emph{\scriptsize{}popage\_dividx} & {\scriptsize{}real between 0 and 1} & \raggedright{}{\scriptsize{}Diversity of age}\tabularnewline 
&&\emph{\scriptsize{}popmale} & {\scriptsize{}percent} & \raggedright{}{\scriptsize{}Fraction of male residential population}\tabularnewline 
&&\emph{\scriptsize{}popstab} & {\scriptsize{}percent} & \raggedright{}{\scriptsize{}Fraction of stable residential population}\tabularnewline 
\cline{3-6}
&&\emph{\scriptsize{}busidens} & {\scriptsize{}businesses/hectare} & \raggedright{}{\scriptsize{}Density of workplaces} & \multirow{6}{*}{\scriptsize{} \shortstack[l]{AGIS: ``Statistik der \\ Unternehmensstruktur \\ (STATENT) 2013 \\auf Hektarbasis''}}\tabularnewline
&&\emph{\scriptsize{}busi\_sec1, busi\_sec2, busi\_sec3} & {\scriptsize{}percent} & \raggedright{}{\scriptsize{}Fraction of businesses in sectors 1, 2 and 3 }\tabularnewline
&&\emph{\scriptsize{}busisec\_dividx} & {\scriptsize{}real between 0 and 1} & \raggedright{}{\scriptsize{}Diversity of workplaces with respect to sector}\tabularnewline
&&\emph{\scriptsize{}emplsec\_dividx} & {\scriptsize{}real between 0 and 1} & \raggedright{}{\scriptsize{}Diversity of employees with respect to sector}\tabularnewline
&&\emph{\scriptsize{}empldens} & {\scriptsize{}employees/hectare} & \raggedright{}{\scriptsize{}Density of employees }\tabularnewline
&&\emph{\scriptsize{}emplmale} & {\scriptsize{}percent} & \raggedright{}{\scriptsize{}Fraction of male employees}\tabularnewline

\cline{2-6}
&\multirow{14}{*}{\rotatebox[origin=c]{90}{\scriptsize{} \shortstack[l]{Crime pattern theory}}}
&\emph{\scriptsize{}land\_indust, land\_park, land\_resi1, ...} & {\scriptsize{}percent} & \raggedright{}{\scriptsize{}Fraction of land usage: industry, park, residential area with 1 story buildings, ...} &\multirow{2}{*}{\scriptsize{} \shortstack[l]{AGIS: ``Bauzonen Schweiz \\(harmonisiert) \\ Ausschnitt AG gemaess MGDM''}} \tabularnewline 
&&\emph{\scriptsize{}land\_dividx } & {\scriptsize{}real between 0 and 1} & \raggedright{}{\scriptsize{}Diversity of land use}\tabularnewline 

\cline{3-6}
&&\emph{\scriptsize{}buildgs\_areafrac} & {\scriptsize{}percent} & \raggedright{}{\scriptsize{}Fraction of area within a grid cell covered by buildings}  &  \multirow{2}{*} {\scriptsize{} \shortstack[l]{AGIS: ``Gebaeude ab \\ Uebersichtsplan 1\_5000''}} \tabularnewline 
&&\emph{\scriptsize{}buildgs\_dens} & {\scriptsize{}buildings/hectare} & {\scriptsize{}Density of buildings within a grid cell}\tabularnewline

\cline{3-6}
&&\emph{\scriptsize{}poi\_infra} & {\scriptsize{}points/hectare} & \raggedright{}{\scriptsize{}Density of infrastructural items such as ATMs, post boxes, waste baskets, etc.} & \multirow{5}{*}{\scriptsize{} \shortstack[l]{OpenStreetMap Switzerland}}  \tabularnewline
&&\emph{\scriptsize{}poi\_shop} & {\scriptsize{}points/hectare} & \raggedright{}{\scriptsize{}Density of shops}\tabularnewline 
&&\emph{\scriptsize{}poi\_public} & {\scriptsize{}points/hectare} & \raggedright{}{\scriptsize{}Density of public buildings such as police stations, hospitals, etc.}\tabularnewline 
&&\emph{\scriptsize{}poi\_edu} & {\scriptsize{}points/hectare} & \raggedright{}{\scriptsize{}Density of educational institutions such as schools, kindergartens, etc.}\tabularnewline 
&&\emph{\scriptsize{}poi\_gastro} & {\scriptsize{}points/hectare} & \raggedright{}{\scriptsize{}Density of gastronomical amenities such as restaurants, bars, etc.}\tabularnewline

\cline{3-6}
&&\emph{\scriptsize{}pub\_hous} & {\scriptsize{}boolean} & \raggedright{}{\scriptsize{}Occurrence of a public housing unit in a grid cell} &  \multirow{1}{*}{\scriptsize{} \shortstack[l]{Bundesamt fuer Wohnungswesen}}  \tabularnewline

\cline{3-6}
&&\emph{\scriptsize{}highway\_exits} & {\scriptsize{}boolean} & \raggedright{}{\scriptsize{}Occurrence of a highway exit within 5,000\,m of the grid cell} & \multirow{4}{*}{\scriptsize{} \shortstack[l]{AGIS: ``Netz Kantons- \\ strassen und Nationalstrassen''}}\tabularnewline
&&\emph{\scriptsize{}border\_cross} & {\scriptsize{}boolean} & \raggedright{}{\scriptsize{}Occurrence of a border crossing within 5,000\,m of the grid cell}\tabularnewline 
&&\emph{\scriptsize{}road\_type} & {\scriptsize{}category} & \raggedright{}{\scriptsize{}Type of a road in the grid cell: from 0 = no road to 3 = high volume road}\tabularnewline
&&\emph{\scriptsize{}intersection} & {\scriptsize{}boolean} & \raggedright{}{\scriptsize{}Occurrence of a major crossroads in the respective or neighboring grid cells}\tabularnewline  

\cline{3-6}
&&\emph{\scriptsize{}pub\_trans} & {\scriptsize{}real between 0 and 1} & \raggedright{}{\scriptsize{}Quality measure for public transport}& \multirow{1}{*}{{\scriptsize{}{AGIS: ``OeV Gueteklassen''}}}\tabularnewline 
\midrule

\multirow{8}{*}{\rotatebox[origin=c]{90}{\scriptsize{} \shortstack[l]{Temporal}}}   &\multirow{8}{*}{\rotatebox[origin=c]{90}{\scriptsize{} \shortstack[l]{Climatic and seasonal}}}  
&\emph{\scriptsize{}dow} & {\scriptsize{}encoded} & \raggedright{}{\scriptsize{}The day of the week} & \multirow{1}{*}{{\scriptsize{}---}} \tabularnewline

\cline{3-6}
&&\emph{\scriptsize{}holiday} & {\scriptsize{}boolean} & \raggedright{}{\scriptsize{}Indicator of whether there is a public holiday} & \multirow{1}{*}{{\scriptsize{}feiertagskalender.ch}} \tabularnewline 

\cline{3-6}
&&\emph{\scriptsize{}temp} & {\scriptsize{}degree celsius} & \raggedright{}{\scriptsize{}Temperature at 12\,am} & \multirow{5}{*}{{\scriptsize{}Darksky API }} \tabularnewline 
&&\emph{\scriptsize{}hum} & {\scriptsize{}percent} & \raggedright{}{\scriptsize{}Humidity at 12\,am}\tabularnewline 
&&\emph{\scriptsize{}discomf} & {\scriptsize{}real} & \raggedright{}{\scriptsize{}Discomfort index}\tabularnewline 
&&\emph{\scriptsize{}daylight} & {\scriptsize{}hours} & \raggedright{}{\scriptsize{}Hours of daylight}\tabularnewline 
&&\emph{\scriptsize{}moon} & {\scriptsize{}float between 0 and 1} & \raggedright{}{\scriptsize{}Moon phase (1 = full moon)}\tabularnewline 

\cline{3-6}
&&\emph{\scriptsize{}event} & {\scriptsize{}integer} & {\scriptsize{}Number of public events on the specific day for the respective cell} & \multirow{1}{*}{{\scriptsize{}events.ch}} \tabularnewline 
\bottomrule
\end{tabular}
}
\caption{{\small{}Theory-informed choice of crime, locational, and temporal features serving as predictors.}}
\label{tab:features}

\end{sidewaystable}

Finally, each outcome crime/non-crime is predicted based on 64 spatio-temporal features grounded in prior criminological research and listed in \Cref{tab:features}. In order account for \emph{(near) repeat victimization}, we craft four features based on recent crime history in the cell and its neighbors. In order to account for \emph{social disorganization theory}, we formulate various socio-demographic, economical, and land-use factors from governmental open data (such as AGIS). For \emph{crime pattern theory}, we exploit other open data platforms (such as OpenStreetMap\footnote{OpenStreetMap Switzerland: \url{https://osm.ch/}. Last accessed: February~28, 2018.}) in order to describe nearby POI and road infrastructure. This amounts to a total of 52 locational features per cell. Lastly, we include eight public features that are predominantly of temporal nature and refer to calendar, weather, and event attributes of the day in each particular cell (such as the Dark Sky API\footnote{Dark Sky API: \url{https://darksky.net/dev}}).

\section{Results}
\label{sec:results}

\subsection{Overall prediction performance}
\label{subsec:performance}

\Cref{tab:final_results} evaluates the different methods to cope with class imbalance introduced previously. The majority class model (which always predicts no crime) does not manage -- by definition -- to predict any hotspot. The second baseline, a random forest with no modification, achieves moderate hit rate and PAI scores. Both over-sampling techniques fail to improve upon these scores, and the first imbalance-aware technique that surpasses the baseline methods in most metrics is the heuristic under-sampling approach. It is followed by the cost-sensitive learning and random under-sampling techniques. 

The best-performing method across all metrics is the hyper-ensemble. It benefits from the fact that it is trained on ten different under-sampled subsets. Through its internal model averaging, its average results outperform those of a single classifier trained on an under-sampled subset. When applied to a coverage area of \SI{5}{\percent}, the hyper-ensemble achieves a hit rate that is \SI{35.9}{\percent} higher than the na{\"i}ve baseline. It even surpasses the best baseline (\ie, random under-sampling) by \SI{5.6}{\percent}. The same pattern is also observed in the PAI scores.

When varying the coverage area, we notice that the hit rate generally improves for the models. Notably, our hyper-ensemble model consistently outperforms all baselines. The most efficient PAI is achieved at the \SI{5}{\percent} level with a value of $4.932$.

We utilize a paired $t$-test in order to assess whether the mean hit rate and PAI results of the hyper-ensemble model are significantly better than the results of the other classifiers. 
 We find that the results of our method are significantly better than the results of the na{\"i}ve baseline at all coverage area levels ($p < 0.001$). Furthermore, the hyper-ensemble also significantly outperforms the random under-sampling approach at \SI{10}{\percent} coverage area ($p < 0.05$), and at \SI{5}{\percent} and \SI{20}{\percent} coverage areas ($p < 0.1$).

\subsection{Sensitivity of prediction to base learners}
We study the sensitivity with regard to different specifications of the base learners and find that the above observations, in terms of models and coverage areas, remain robust. \Cref{tab:baselearners_results} lists the performance of the two best model specifications (random under-sampling and hyper-ensemble of random under-sampling) with the four different base learners introduced in the previous section. We notice relatively similar performance results, whereby the random forest setup wins in most configurations, followed closely by the AdaBoost setup. 

\subsection{Sensitivity of prediction by population density}
\label{subsec:geo_analysis}
We study the sensitivity of the prediction performance across different levels of population density in order to understand whether there are specific regions where the model performs better (or worse). For this purpose, we split the dataset according to percentiles of the population density attribute in order to have three separate datasets with a similar number of samples: low population density ($\text{\emph{popdens}} < 2.25 \text{\ residents/hectare}$), medium population density ($2.25\text{\ residents/hectare} \le \text{\emph{popdens}} \le 16.75 \text{\ residents/hectare}$), and  high population density ($\text{\emph{popdens}} > 16.75 \text{\ residents/hectare}$). Most crime occurs in the high population density areas (\ie, 1,335 incidents), some in areas of medium population density (\ie, 398 incidents), and considerably less in low population density regions (\ie, 151 incidents).

For each subset, a model was trained analogous to the previous experiment. The performance results are listed in \Cref{tab:subset_results}, revealing the following key finding: the higher the population density, the better the models are at ranking higher risk areas. Still, the overall model, having access to all data, achieves the best hit rates and PAI scores. For example, the hit rate score in the overall model outperforms the hit rate in the less populated areas by as much as $134.3\%$ and the hit rate in the more populated areas by $64.0\%$.

\begin{sidewaystable}
\begin{table}[H]
\centering
\begin{tabular}{lSSSSSSS}
\toprule
\textbf{\footnotesize{}Classifier} 
&\multicolumn{2}{c}{\textbf{\footnotesize{}5\,\% Coverage area}} 
&\multicolumn{2}{c}{\textbf{\footnotesize{}10\,\% Coverage area}} 
&\multicolumn{2}{c}{\textbf{\footnotesize{}20\,\% Coverage area}}
&\textbf{\footnotesize{}AUC}\\

\cmidrule(lr){2-3}
\cmidrule(lr){4-5}
\cmidrule(lr){6-7}
  
&\textbf{\footnotesize{}Hit rate} &\textbf{\footnotesize{}PAI} 
&\textbf{\footnotesize{}Hit rate} &\textbf{\footnotesize{}PAI} 
&\textbf{\footnotesize{}Hit rate} &\textbf{\footnotesize{}PAI}
&\\
\midrule

{\scriptsize{}Majority class classifier} &{\scriptsize{0.0\,\%}} &{\scriptsize{}0.000} &{\scriptsize{}0.0\,\%} &{\scriptsize{}0.000} &{\scriptsize{}0.0\,\%} &{\scriptsize{}0.000} &{\scriptsize{}0.000}\\

{\scriptsize{}Na{\"i}ve classifier} &{\scriptsize{18.1\,\%}} &{\scriptsize{}3.621} &{\scriptsize{}32.5\,\%} &{\scriptsize{}3.249} &{\scriptsize{}53.0\,\%} &{\scriptsize{}2.651} &{\scriptsize{}0.754}\\
\midrule

{\scriptsize{}Cost-sensitive learning} &{\scriptsize{}21.0\,\%} &{\scriptsize{}4.186} &{\scriptsize{}36.1\,\%} &{\scriptsize{}3.611} &{\scriptsize{}58.7\,\%} &{\scriptsize{}2.935} &\scriptsize{}0.769\\

{\scriptsize{}Random over-sampling} &{\scriptsize{}15.5\,\%} &{\scriptsize{}3.094} &{\scriptsize{}25.2\,\%} &{\scriptsize{}2.523} &{\scriptsize{}40.8\,\%} &{\scriptsize{}2.040} &{\scriptsize{}0.593}\\

{\scriptsize{}Random under-sampling} &{\scriptsize{}23.3\,\%} &{\scriptsize{}4.665} &{\scriptsize{}38.4\,\%} &{\scriptsize{}3.840} &{\scriptsize{}59.1\,\%} &{\scriptsize{}2.953} &{\scriptsize{}0.773}\\

{\scriptsize{}Heuristic over-sampling} &{\scriptsize{}16.2\,\%} &{\scriptsize{}3.252} &{\scriptsize{}26.5\,\%} &{\scriptsize{}2.649} &{\scriptsize{}41.3\,\%} &{\scriptsize{}2.063} &{\scriptsize{}0.590}\\

{\scriptsize{}Heuristic under-sampling} &{\scriptsize{}12.5\,\%} &{\scriptsize{}2.495} &{\scriptsize{}21.8\,\%} &{\scriptsize{}2.181} &{\scriptsize{}43.2\,\%} &{\scriptsize{} 2.159} &{\scriptsize{}0.721}\\
\midrule

{\scriptsize{}Hyper-ensemble} &{\scriptsize{}\textbf{24.6\,\%}} &{\scriptsize{}\textbf{4.932}} &{\scriptsize{}\textbf{40.2\,\%}} &{\scriptsize{}\textbf{4.020}} &{\scriptsize{}\textbf{60.4\,\%}} &{\scriptsize{}\textbf{3.021}} &{\scriptsize{}\textbf{0.779}}\\


\bottomrule
\end{tabular}
\caption{Test set performance comparison of the machine learning methods proposed to cope with class imbalance. All models are trained using random forests as base learners.}
\label{tab:final_results}
\end{table}

\begin{table}[H]
\centering
\begin{tabular}{llSSSSSSS}
\toprule
\textbf{\footnotesize{}Classifier} &\textbf{\footnotesize{}Base learner} 
&\multicolumn{2}{c}{\textbf{\footnotesize{}5\,\% Coverage area}} 
&\multicolumn{2}{c}{\textbf{\footnotesize{}10\,\% Coverage area}} 
&\multicolumn{2}{c}{\textbf{\footnotesize{}20\,\% Coverage area}}
&\textbf{\footnotesize{}AUC}\\
\cmidrule(lr){3-4}
\cmidrule(lr){5-6}
\cmidrule(lr){7-8}
&  
&\textbf{\footnotesize{}Hit rate} &\textbf{\footnotesize{}PAI} 
&\textbf{\footnotesize{}Hit rate} &\textbf{\footnotesize{}PAI} 
&\textbf{\footnotesize{}Hit rate} &\textbf{\footnotesize{}PAI}
&\\
\midrule

{\scriptsize{}Random under-sampling} &{\scriptsize{}Random forest}  &{\scriptsize{}23.3\,\%} &{\scriptsize{}4.665} &{\scriptsize{}38.4\,\%} &{\scriptsize{}3.840} &{\scriptsize{}59.1\,\%} &{\scriptsize{}2.952} &{\scriptsize{}0.773}\\
{\scriptsize{}Random under-sampling} &{\scriptsize{}AdaBoost} &{\scriptsize{}23.2\,\%} &{\scriptsize{}4.654} &{\scriptsize{}37.7\,\%} &{\scriptsize{}3.771} &{\scriptsize{}58.7\,\%} &{\scriptsize{}2.933} &{\scriptsize{}0.771}\\
{\scriptsize{}Random under-sampling} &{\scriptsize{}L2 logistic regression} &{\scriptsize{}24.6\,\%} &{\scriptsize{}4.922} &{\scriptsize{}37.1\,\%} &{\scriptsize{}3.712} &{\scriptsize{}57.3\,\%} &{\scriptsize{}2.866} &{\scriptsize{}0.768}\\
{\scriptsize{}Random under-sampling} &{\scriptsize{}L1 logistic regression} &{\scriptsize{}21.8\,\%} &{\scriptsize{}4.363} &{\scriptsize{}34.1\,\%} &{\scriptsize{} 3.407} &{\scriptsize{}54.4\,\%} &{\scriptsize{}2.717} &{\scriptsize{}0.758}\\

\midrule
{\scriptsize{}Hyper-ensemble} &{\scriptsize{}Random forest}
&{\scriptsize{}24.6\,\%} &{\scriptsize{}4.932} &{\scriptsize{}\textbf{40.2\,\%}} &{\scriptsize{}\textbf{4.020}} &{\scriptsize{}\textbf{60.4\,\%}} &{\scriptsize{}\textbf{3.021}} &{\scriptsize{}\textbf{0.779}}\\

{\scriptsize{}Hyper-ensemble} &{\scriptsize{}AdaBoost} 
&{\scriptsize{}23.3\,\%} &{\scriptsize{}4.659} 
&{\scriptsize{}38.1\,\%} &{\scriptsize{}3.809} &{\scriptsize{}59.2\,\%} &{\scriptsize{}2.959} &{\scriptsize{}0.772}\\

{\scriptsize{}Hyper-ensemble} &{\scriptsize{}L2 logistic regression}  
&{\scriptsize{}\textbf{24.7\,\%}} &{\scriptsize{}\textbf{4.936}}
&{\scriptsize{}37.3\,\%} &{\scriptsize{}3.731} &{\scriptsize{}57.4\,\%} &{\scriptsize{}2.868} &{\scriptsize{}0.769}\\

{\scriptsize{}Hyper-ensemble} &{\scriptsize{}L1 logistic regression} 
&{\scriptsize{}22.1\,\%} &{\scriptsize{}4.414} 
&{\scriptsize{}34.1\,\%} &{\scriptsize{}3.409} &{\scriptsize{}54.5\,\%} &{\scriptsize{}2.722} &{\scriptsize{}0.758}\\
\bottomrule
\end{tabular}
\caption{Test set performance comparison of different base learners reveals only marginal sensitivity to the choice of the base learner.}
\label{tab:baselearners_results}
\end{table}
\end{sidewaystable}

\begin{sidewaystable}
\begin{table}[H]
\centering
\begin{tabular}{llSSSSSSS}
\toprule
\textbf{\footnotesize{}{Population density}} &\textbf{\footnotesize{}Feature set} 
&\multicolumn{2}{c}{\textbf{\footnotesize{}5\,\% Coverage area}} 
&\multicolumn{2}{c}{\textbf{\footnotesize{}10\,\% Coverage area}} 
&\multicolumn{2}{c}{\textbf{\footnotesize{}20\,\% Coverage area}}
&\textbf{\footnotesize{}AUC}\\
\cmidrule(lr){3-4}
\cmidrule(lr){5-6}
\cmidrule(lr){7-8}
&  
&\textbf{\footnotesize{}Hit rate} &\textbf{\footnotesize{}PAI} 
&\textbf{\footnotesize{}Hit rate} &\textbf{\footnotesize{}PAI} 
&\textbf{\footnotesize{}Hit rate} &\textbf{\footnotesize{}PAI}
&\\
\midrule

\multirow{4}{*}{\scriptsize{}All} 
&{\scriptsize{}Crime}     &{\scriptsize{}16.1\,\%} &{\scriptsize{}3.213}
&{\scriptsize{}27.4\,\%} &{\scriptsize{}2.743} &{\scriptsize{}39.0\,\%} &{\scriptsize{}1.949} &{\scriptsize{}0.581}\\
&{\scriptsize{}Temporal}  &{\scriptsize{}6.5\,\%} &{\scriptsize{}1.292}
&{\scriptsize{}12.5\,\%} &{\scriptsize{}1.255} &{\scriptsize{}21.1\,\%} &{\scriptsize{}1.057} &{\scriptsize{}0.498}\\
&{\scriptsize{}Spatial}   &{\scriptsize{}23.3\,\%} &{\scriptsize{}4.655}
&{\scriptsize{}37.1\,\%} &{\scriptsize{}3.711} &{\scriptsize{}57.2\,\%} &{\scriptsize{}2.860} &{\scriptsize{}0.771}\\

&{\scriptsize{All}} 
&{\scriptsize{}24.6\,\%} &{\scriptsize{}4.932}
&{\scriptsize{}40.2\,\%} &{\scriptsize{}4.020} &{\scriptsize{}60.4\,\%} &{\scriptsize{}3.021} &{\scriptsize{}0.779}\\
\midrule

\multirow{4}{*}{\scriptsize{}Low population density}
&{\scriptsize{}Crime}   &{\scriptsize{}4.0\,\%} &{\scriptsize{}0.813}
&{\scriptsize{}5.0\,\%} &{\scriptsize{}0.508} &{\scriptsize{}8.0\,\%} &{\scriptsize{}0.402} &{\scriptsize{}0.157}\\
&{\scriptsize{}Temporal}&{\scriptsize{}0.7\,\%} &{\scriptsize{}0.142} 
&{\scriptsize{}3.8\,\%} &{\scriptsize{}0.385} &{\scriptsize{}6.5\,\%} &{\scriptsize{}0.324} &{\scriptsize{}0.154}\\
&{\scriptsize{}Spatial} &{\scriptsize{}10.5\,\%} &{\scriptsize{}2.096}
&{\scriptsize{}15.7\,\%} &{\scriptsize{}1.568} &{\scriptsize{}21.2\,\%} &{\scriptsize{}1.060} &{\scriptsize{}0.251}\\

&{\scriptsize{All}} 
&{\scriptsize{}10.5\,\%} &{\scriptsize{}2.096}
&{\scriptsize{}15.7\,\%} &{\scriptsize{}1.568} &{\scriptsize{}21.2\,\%} &{\scriptsize{}1.060} &{\scriptsize{}0.251}\\
\midrule

\multirow{4}{*}{\scriptsize{}Medium population density}
&{\scriptsize{}Crime}   &{\scriptsize{}8.4\,\%} &{\scriptsize{}1.686}
&{\scriptsize{}13.6\,\%} &{\scriptsize{}1.357} &{\scriptsize{}20.4\,\%} &{\scriptsize{}1.019} &{\scriptsize{}0.355}\\
&{\scriptsize{}Temporal}&{\scriptsize{}3.1\,\%} &{\scriptsize{}0.623}
&{\scriptsize{}7.6\,\%} &{\scriptsize{}0.763} &{\scriptsize{}14.5\,\%} &{\scriptsize{}0.726} &{\scriptsize{}0.327}\\
&{\scriptsize{}Spatial} &{\scriptsize{}14.0\,\%} &{\scriptsize{}2.805}
&{\scriptsize{}20.4\,\%} &{\scriptsize{}2.043} &{\scriptsize{}31.5\,\%} &{\scriptsize{}1.572} &{\scriptsize{}0.447}\\

&{\scriptsize{All}} 
&{\scriptsize{}15.6\,\%} &{\scriptsize{}3.123}
&{\scriptsize{}21.5\,\%} &{\scriptsize{}2.156} &{\scriptsize{}29.9\,\%} &{\scriptsize{}1.493} &{\scriptsize{}0.448}\\
\midrule
 
\multirow{4}{*}{\scriptsize{}High population density} 
&{\scriptsize{}Crime}   &{\scriptsize{}11.6\,\%} &{\scriptsize{}2.320}
&{\scriptsize{}21.9\,\%} &{\scriptsize{}2.185} &{\scriptsize{}35.1\,\%} &{\scriptsize{}1.756} &{\scriptsize{}0.561}\\ 
&{\scriptsize{}Temporal}&{\scriptsize{}6.0\,\%} &{\scriptsize{}1.199}
&{\scriptsize{}11.3\,\%} &{\scriptsize{}1.132} &{\scriptsize{}21.7\,\%} &{\scriptsize{}1.083} &{\scriptsize{}0.486}\\
&{\scriptsize{}Spatial} &{\scriptsize{}15.1\,\%} &{\scriptsize{}3.033}
&{\scriptsize{}23.5\,\%} &{\scriptsize{}2.348} &{\scriptsize{}39.2\,\%} &{\scriptsize{}1.960} &{\scriptsize{}0.633}\\

&{\scriptsize{All}} 
&{\scriptsize{}15.0\,\%} &{\scriptsize{}3.013}
&{\scriptsize{}27.2\,\%} &{\scriptsize{}2.721} &{\scriptsize{}41.8\,\%} &{\scriptsize{}2.091} &{\scriptsize{}0.651}\\
\bottomrule
 
\end{tabular}
\caption{Test set performance comparison of crime, temporal, spatial, and all features across different levels of population density. All models are trained using a hyper-ensemble of random under-sampling with random forests.} 
\label{tab:subset_results}
\end{table}
\end{sidewaystable}



\subsection{Sensitivity of prediction to temporal resolution}

We also study the sensitivity with regard to the temporal resolution, by processing the target variable and the features at weekly granularity. This yields a dataset of 1,552,797 spatio-temporal observations (= 10,149 grid cells $\times$ 153 weeks) with a pronounced class imbalance of $0.56\%$ instances belonging to the positive class (\ie, crime). In this scenario, the winning hyper-ensemble of random under-sampling based on random forests achieves a total AUC of \SI{0.844}. This corresponds to a hit rate of \SI{51.9}{\percent} at \SI{5}{\percent} coverage level, and to a hit rate of \SI{74.4}{\percent} at \SI{20}{\percent} coverage level. As such, weekly burglary hotspots prove to be easier to predict as daily burglary hotspots. Yet, daily predictions are desired for tactical decision-making in law enforcement.

\subsection{Relevance of feature sets}
\label{subsec:feature_analysis}

Finally, we perform a direct comparison of the different predictors in terms of their predictive power for predictive policing in areas with low population density. We analyze how the prediction performance varies across three subsets of all candidate features: crime features (recent crime history in the area and surroundings), locational features (socio-demographic, land-use, POI and infrastructure factors), and primarily temporal features (calendar, weather, and event attributes). Per \Cref{tab:subset_results}, we conclude that, in all areas, the locational features yield the best predictive results, approaching the results of the models with all features. This highlights the overall importance of both social disorganization theory and crime pattern theory when modeling crime patterns. Second is the model with features inferred from the recent crime incidents in the area, as per insights from (near) repeat victimization. The least predictive features turn out to be the temporal features. For instance, when training on samples spanning all population density levels, the model utilizing the complete feature set outperforms the temporal-only setting by \SI{278.5}{\percent} and the crime-only setting by \SI{52.8}{\percent}. These results refer to the hit rate at \SI{5}{\percent} coverage level, but, as depicted in \Cref{fig:hitrate_coverage_features}, the large improvement of the full model against the crime-only baseline holds at all coverage levels. 

\begin{figure}[H]
\centerline{\includegraphics[width=0.5\textwidth]{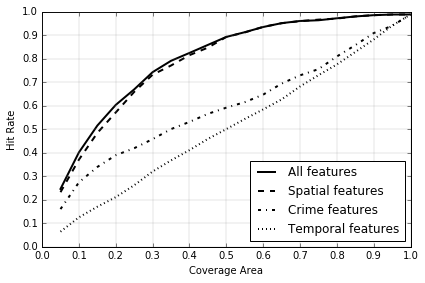}}
\caption{Surveillance plot: hit rate as a function of coverage area  when considering different feature sets.}
\label{fig:hitrate_coverage_features}
\end{figure}

When comparing the different levels, we observe that, relative to all other zones, the spatial features perform best in the low-density areas, whereas the past crime features perform best in the high-density areas. 

\section{Discussion}
\label{sec:discussion}

\subsection{Interpretation and link to the literature}
This paper establishes the effectiveness of tailored machine learning in terms of identifying places and times of elevated burglary risk in a wide and heterogeneous region with low population density. The superior performance of the proposed hyper-ensemble can be attributed to two main factors: (1)~an innovative strategy that deals with the strong class imbalance by fusing the concepts of under-sampling and ensemble learning, and (2)~the incorporation of state-of-the-art spatio-temporal predictors of crime into the model.

The results achieved by our approach in a \emph{sparsely-populated} environment are -- despite the methodological challenges -- on a par with or better than the latest results achieved in \emph{densely-populated} environments, which, to their advantage, do not suffer from severe class imbalance. As an illustration, \citet{Adepeju2016} report a hit rate of \SI{51.5}{\percent}, \ie, a PAI of $2.57$, for burglary in South Chicago, while \citet{Gerber2014} reports a hit rate of \SI{45.0}{\percent}, \ie, a PAI of $2.25$, for crime in Los Angeles, both at \SI{20}{\percent} coverage area. Furthermore, \citet{Rummens2017} report the optimal values of \SI{25.1}{\percent} for hit rate and $3.95$ for PAI, which were achieved at \SI{20}{\percent} cut-off probability for home burglary in a Belgian city (this translates to approximately \SI{6}{\percent} coverage area). Hence, our hit rate of \SI{60.4}{\percent} at \SI{20}{\percent} coverage area attains a similar performance, despite the more challenging setting of sparsity.

We further investigate whether the predictive performance varies across different levels of population density. We find that the presented system performs better in regions with highest levels of population density. This is not surprising, since such areas: (1)~experience more crime and thus provide the algorithm with more training examples of the positive class, and (2)~coincide with a wider distribution of the features, which allows the algorithm to discover more discriminative patterns.

As crime in less populated areas might not be equally affected by the different spatio-temporal factors considered, we discuss the prognostic capacity of the separate crime, spatial, and temporal feature sets. When looking at the whole study area, we find that the high number of spatial features inspired by the social disorganization and crime pattern theories predict crime best. These are followed by recent crime history features crafted according to the (near) repeat victimization phenomena, while a model trained on temporal features delivers only marginal improvement over the na{\"i}ve baseline. Most importantly, the additional features describing the environment and time considerably improve the predictive power of the models relying only on crime features.

Different patterns emerge when comparing different levels of population density. In that case, the importance of past crime features increases with increasing levels of density. For instance, in less populated areas, the model based on locational features based o social disorganization theory and crime pattern theory achieves the same results as the model trained on all features, while the performance of the model based on crime features is inferior by a factor of about $2.5$. For more populated areas, the ratio shrinks to about $1.3$. his tendency is consistent with the current literature and practices\footnote{PredPol: \url{https://www.predpol.com/}. Last accessed: March~11, 2018.\label{note1}} in urban predictive policing, which rely heavily on insights derived from the notion of (near) repeat victimization.  

\subsection{Implications for management and academia}
A number of studies, including randomized controlled trials, have revealed the operative potential of short-term crime prediction models \citep{ Chainey2008, Mohler2015}. In response, predictive policing solutions, such as PredPol\textsuperscript{\ref{note1}} and HunchLab\footnote{HunchLab: \url{https://www.hunchlab.com/}. Last accessed: March~11, 2018.}, have been developed and integrated into the daily work of police officers in several major cities. This has led the National Research Council Committee to review research on police policies and practices, finding strong evidence that taking a focused geographic approach to the problem of crime can increase the effectiveness of policing and could lead to better public decision-making \citep{NationalResearchCouncil2004}. This renders our work timely and highly relevant. We contribute to this effort by studying areas with low population density that have been overlooked in previous research due to the inherent challenges of this undertaking.


Advances in big data and machine learning over the past decade have enabled the implementation of location and non-location analytics in decision support research \citep{Pick2017} by utilizing proprietary and open data for different applications such as optimizing car-sharing \citep{Barann2017,Willing2017}, predicting crimes \citep{Gerber2014,Vomfell2018}, or preventing traffic accidents \citep{Ryder2017}. Spatially-referenced data that is usually inaccessible has been made available to academic researchers, allowing us to advance the literature on location analytics for spatial decision support in smart cities \citep{Pick2017_smart_cities}, with an emphasis on under-researched regions with low population density. 
 Furthermore, our study illustrates the successful integration of several existing theories and their innovative application to public decision-making. 

The tandem of a constant increase in the ubiquity of data, in combination with exponentially advancing computational power, has led to numerous applications of big data analytics in practice. Based on the lessons learned from the digitalization of the private sector, stakeholders in the public sector are now learning how to adapt their processes and services for the 21st century -- with police forces representing one example \citep{Taylor2015}. Hence, the proposed methodology can be directly leveraged by decision-makers in law enforcement, both private (\eg, private security firms) and especially public (\eg, police forces). We provide a fully-fledged approach that can be integrated as a prediction module in a spatial decision support system, as the one presented in \Cref{fig:decision_support}. Based on the current available patrolling resources, a decision-maker can set the maximum coverage area and identify the top crime hotspots.

In the broader sense and inspired by our work, similar approaches utilizing imbalance-aware machine learning techniques can be envisioned to forecast other sparse spatio-temporal phenomena such as demand for prescriptions, ambulance calls, or 311/911 calls. 

\subsection{Limitations and potential for future research}

With the goal of developing a crime prediction approach for areas with low density population, we have leveraged random forests in a hyper-ensemble approach utilizing a super-set of features that have their origins in criminological theory and empirical studies. Although random forests benefit from strong performance in this predictive setup, the resulting decision rules are largely data-driven and thus lack the same interpretability as theory. Hence, they are not the ideal instrument for testing the theoretical contribution of each individual feature -- carefully crafted explanatory models and randomized controlled trials would be proper instruments if that objective was desired.

Future research could explore additional methods for highly imbalanced classifications in the domain of unsupervised learning, such as novelty detection and outlier detection. For novelty detection, the training data is not polluted by outliers, and the interest is on detecting anomalies in new observations. 

Another potential avenue for future work consists of hourly crime prediction. The current limitation lies in the fact that burglaries are often discovered only post hoc. Because of that, the actual time of the burglary is unknown and can merely be approximated. With more effort on precise reporting, hourly analyses could further bolster decision support.

While we already worked with an extensive set of features that adheres to best-practice recommendations from prior research, one could potentially investigate the prognostic capacity of further features, such as, \eg, mobility or social media data, in a low population density setting. The current limitation for these datasets is their extreme sparsity in areas with low population density, making their application unfeasible at the moment. Though with the proliferation of location-based services, this might be worth revisiting in the future. In addition to expanding the feature set, it would be interesting to evaluate the presented features with respect to other types of crimes, such as theft or assaults.

\section{Conclusion}
\label{sec:conclusion}
The aim of this work was to explore the potential of predictive policing in a low population density setting. While areas with high population density have been subject to extensive research, decision support for law enforcement that is designed for less populated areas and their unique characteristics is scarce.
Our results have demonstrated the successful application of imbalance-aware machine learning techniques to the task of building a ranking system that identifies regions and times of elevated burglary risk in a large, heterogeneous area. One of the major challenges in the prediction setup is the sparse nature of crime, which is commonly referred to as a severe class imbalance. In our study, it is coped with by proposing a hyper-ensemble that combines under-sampling and ensemble learning in an effective strategy. 
This approach outperforms traditional techniques based on (heuristic) under- and over-sampling, as well as cost-sensitive learning. By incorporating various spatio-temporal crime-correlating factors into the model, we significantly improve upon the predictive performance of models that rely solely on past crime data. When dividing the study area according to three different levels of population density, we notice an increase in predictive performance in areas with higher population densities. We conclude that the proposed approach can be useful for areas with low population density, which up to now have been neglected within the domain of crime forecasting. 




\section*{References}

{\sloppy
\setlength{\bibsep}{3pt}
\bibliographystyle{model5-names}
\bibliography{dss_paper}

}







\end{document}